\documentclass[floatfix,aps,amsmath,amssymb,superscriptaddress,twocolumn]{revtex4-2}
\usepackage{bm,bbm}
\usepackage{amssymb,amsfonts,amsmath}
\usepackage{graphicx}
\usepackage{xcolor}
\usepackage{ulem}
\usepackage{SIunits}
\usepackage[utf8]{inputenc}
\usepackage[english]{babel}
\usepackage{hyperref}
\usepackage{ulem}
\usepackage{braket}
\usepackage{bpchem}
\usepackage{tikz}
\usepackage{pdfpages}
\usepackage{url}
\graphicspath{{figures/}}
\makeatletter
\AtBeginDocument{\let\LS@rot\@undefined}
\makeatother

\newcommand{\rs}[1]{\rm{\scriptscriptstyle #1}}
\newcommand{\RM}[1]{\MakeUppercase{\romannumeral #1}}
\newcommand{\lp}{l_\perp}

\allowdisplaybreaks

\begin{document}

%%%%%%%%%%%%%%%%%%%% Title %%%%%%%%%%%%%%%%%%%% 
%
\title{
	Ground-state stability and excitation spectrum of a one-dimensional dipolar supersolid
}

%%%%%%%%%%%%%%%%%%%% Authors %%%%%%%%%%%%%%%%%%%% 
\author{Tobias Ilg}
\affiliation{
  Institute for Theoretical Physics \RM3 and Center for Integrated Quantum Science and Technology, University of Stuttgart, DE-70550 Stuttgart, Germany
}
\author{Hans Peter B\"uchler}
\affiliation{
  Institute for Theoretical Physics \RM3 and Center for Integrated Quantum Science and Technology, University of Stuttgart, DE-70550 Stuttgart, Germany
}

%%%%%%%%%%%%%%%%%%%% Date %%%%%%%%%%%%%%%%%%%%
\date{\today}

%%%%%%%%%%%%%%%%%%%% Abstract %%%%%%%%%%%%%%%%%%%%
\begin{abstract}
	We study the behavior of the excitation spectrum across the quantum phase transition from a superfluid to a supersolid phase of a dipolar Bose gas confined to a one-dimensional geometry.
	Including the leading beyond-mean-field effects within an effective Hamiltonian, 
	the analysis is based on Bogoliubov theory with several order parameters accounting for the superfluid as well as  solid structure. 
	We find fast convergence of the ground-state energy in the supersolid with the number of order parameters
	and demonstrate a stable excitation spectrum with two Goldstone modes and an amplitude mode in the low-energy regime. 
	Our results suggest that there exists an experimentally achievable parameter regime for dysprosium atoms, 
	where the supersolid phase exhibits a stable excitation spectrum in the thermodynamic limit 
	and the transition into the supersolid phase is of second order driven by the roton instability.
\end{abstract}

\maketitle

%%%%%%%%%%%%%%%%%%%% Introduction %%%%%%%%%%%%%%%%%%%%
\section{Introduction}
Breakthrough experiments with weakly interacting dipolar Bose gases have recently demonstrated the appearance of a supersolid phase in elongated traps \cite{Tanzi2019a,Boettcher2019,Chomaz2019}. 
The phase transition is accompanied by the appearance of a roton minimum in the excitation spectrum  \cite{Santos2003,Petter2019}. 
Remarkably, the description within the Gross-Pitaevskii formalism requires the inclusion of the leading beyond-mean-field correction for the stability of the droplets  
\cite{FerrierBarbut2016,Petrov2015}. 
Such numerical studies within the experimental three-dimensional setting are in good agreement with experimental observations in dipolar quantum gases \cite{Boettcher2020,Chomaz2023}. 
In this paper, we study whether the supersolid phase exhibits stable excitations in thermodynamic limit
by deriving the  low-energy excitation spectrum across the phase transition from the superfluid to the supersolid phase within Bogoliubov theory.

The possibility  of a ground state  for interacting bosonic particles which combines the density modulation of a solid and the frictionless flow of a 
superfluid has been shown by  Leggett  \cite{Leggett1970}.  
Especially, solid helium has been discussed as a candidate for this exotic state of matter, a system with nearly one atom per unit cell \cite{Balibar2010,Boninsegni2012,Chan2013}.
In contrast, the current experiments with dysprosium atoms  work in a rather complementary regime and realize a supersolid state with several thousand atoms on each lattice site: 
a parameter regime, where one can expect mean-field theory to describe the supersolid phase accurately.
The main ingredient for the appearance of the supersolid state is the combination of a tunable short-range interaction with a magnetic dipole-dipole interaction.  
For increasing influence of the dipolar interaction, such systems can undergo an instability towards the formation of quantum droplets 
\cite{Kadau2016,FerrierBarbut2016,FerrierBarbut2016a,Chomaz2016,Wenzel2017,FerrierBarbut2018,Waechtler2016,Bisset2016,Macia2016,Saito2016,Cinti2017}, 
as well as self-bound droplets \cite{Schmitt2016,Boettcher2019a,Waechtler2016,Baillie2016,Baillie2017,Cinti2017a}, 
or supersolid states 
\cite{Tanzi2019a,Boettcher2019,Chomaz2019,Guo2019,Tanzi2019,Natale2019,Tanzi2021,Sohmen2021,Petter2021,Biagioni2022,Bland2022,Norcia2022,Roccuzzo2019,Hertkorn2019,Zhang2019,Roccuzzo2020,Gallemi2020,Blakie2020a,Zhang2021,Ancilotto2021,Hertkorn2021a,Hertkorn2021,Poli2021,Roccuzzo2022,SanchezBaena2022,Buehler2022,Smith2023,Sindik2022,Bombin2017,Bombin2019}. 
An important observation was that these states are only stabilized by the leading beyond-mean-field correction, 
which provides an additional contribution  to the energy functional stabilizing the system at higher densities  against a collapse \cite{FerrierBarbut2016}; 
such a stabilization has  previously been predicted for Bose mixtures \cite{Petrov2015} and later also experimentally observed \cite{Cabrera2017,Semeghini2018}. 
Within local-density approximation, this additional term can be included into the Gross-Pitaevskii functional 
and forms the basis for extensive numerical studies of the supersolid state and its excitation spectrum.
However, the nature of the transition is often difficult to access in such fully numerical approaches in a finite-size setting \cite{Roccuzzo2019,Smith2023}. 

Here, we present an analytical study on the nature of the quantum phase transition from a superfluid to the supersolid in the thermodynamic limit 
and analyze the excitation spectrum across the transition. 
The analysis is based on Bogoliubov theory in a one-dimensional setting, where we account for the transverse confinement by a variational
ansatz and include the beyond-mean-field contributions within an effective Hamiltonian. 
The supersolid state is described by the macroscopic occupation of additional modes, each mode contributing a higher harmonic to the modulated ground-state wave function.
The influence of each harmonic is characterized by a new order parameter.
We demonstrate that the ground-state energy close to the quantum phase transition converges very quickly for an increasing
number of order parameters, and demonstrate that the excitation spectrum is stable. Especially, we find in the low-energy regime two gapless modes in agreement 
with the two broken continuous symmetries as well as a gapped amplitude mode for the solid structure. 
For parameters comparable to current experiments with dysprosium atoms, 
this analysis confirms that the beyond-mean-field corrections stabilize the supersolid phase within a one-dimensional geometry and demonstrates that the phase 
transition within mean-field theory can be of second order and driven by the roton instability, depending on exact system parameters. 
 
The paper is structured as follows. 
In Sec.~\ref{sec:superfluidphase} we discuss the treatment of the superfluid phase in the one-dimensional geometry and how to include quantum fluctuations in our approach.
Within our model, we discuss excitations in the superfluid in Sec.~\ref{sec:excitationsuperfluid} and briefly examine the roton instability in Sec.~\ref{sec:rotoninstability}.
Then, we adapt our approach in Sec.~\ref{sec:groundstatess} to describe the one-dimensional  supersolid  and calculate its excitation spectrum in Sec.\ref{sec:excitationsss}.
Lastly, we summarize our results in Sec.~\ref{sec:conclusion}.

%%%%%%%%%%%%%%%%%%%% Superfluid phase %%%%%%%%%%%%%%%%%%%%
\section{Superfluid phase}
\label{sec:superfluidphase}

In this paper, we calculate the excitation spectrum across the phase transition from a Bose-Einstein condensate into the supersolid regime 
within a simple reduced three-dimensional model \cite{Blakie2020a} 
using Bogoliubov theory and  include beyond-mean-field corrections in local-density approximation.
The model has recently been shown to produce qualitatively accurate predictions \cite{Smith2023}.
We  consider a gas of trapped dipolar bosons with mass $m$, which are tightly confined in the $x$-$y$ plane by a harmonic confinement but free along the $z$ direction. 
The validity of the local-density approximation requires that the characteristic healing length $\xi$ of the superfluid 
is smaller than the harmonic oscillator length $l_{\perp}$ of the transverse confinement, i.e., $\xi/l_{\perp} \ll 1$  (see Ref.~\cite{Ilg2018}). 
In the following, we are interested in the low-energy excitations and the stability analysis of the supersolid phase. 
For these considerations, transverse excitations can be ignored 
and we make a variational ansatz for the transverse wave function, $\psi(x,y)=\exp[-(\nu x^2 + y^2 /\nu)/2(\sigma\lp)^2]/(\sqrt{\pi}\sigma\lp)$, 
with the dimensionless variational parameters $\sigma$ and $\nu$ determined by minimizing the ground-state energy (see below). 
Within this variational framework, the microscopic Hamiltonian becomes one dimensional and consists of two parts, $H_0+H_{\rs{I}}$.
The single-particle Hamiltonian is described by $H_0=\sum_q [\epsilon_0(q)+ E_{t}(\sigma,\nu)] a^\dagger_q a_q$,
where $a_{q}^{\dagger}$ ($a_{q}$) are the bosonic creation (annihilation) operators of particles with momentum $q$, respectively.
The first term accounts for the kinetic energy along the tube with the dispersion relation  $\epsilon_{0}(q)=\frac{\hbar^2q^2}{2m}$, 
while $E_{t}(\sigma,\nu)=E_{\perp}(\frac{1}{\nu}+\nu)(\frac{1}{\sigma^2}+\sigma^2)/4$ accounts for the energy of the particles in the transverse trap with $E_{\perp}=\hbar^2/m \lp^2$.
The particles interact via a short-range contact interaction characterized by the $s$-wave scattering length $a_{\rs{s}}$ and the anisotropic magnetic dipole-dipole interaction  with 
strength $a_{\rs{dd}}$. The dipoles are aligned perpendicular to the $z$ direction.
By integrating out the transverse degrees of freedom using the variational wave function, the two-body potential in momentum space is well described by  \cite{Blakie2020}
\begin{equation}
	V(q)=
	g_{\rs{1D}}
  	\left[ 
    		1+\varepsilon_{\rs{dd}}
      		\left( 
			\frac{3[1-Q e^Q \Gamma(0,Q)]}{1+\nu} -1
      		\right)
    	\right],
  \nonumber
\end{equation}
with $g_{\rs{1D}}=2\hbar^2a_{\rs{s}}/(m\sigma^2\lp^2)$, $Q=\sqrt{\nu}(q\sigma\lp)^2/2$, and $\varepsilon_{\rs{dd}}=a_{\rs{dd}}/a_{\rs{s}}$. 
Here,  $\Gamma(s,x)$  denotes the incomplete gamma function.
Corrections of $g_{\rs{1D}}$ due to the confinement-induced resonance are only relevant for $\xi/l_{\perp} \gg 1$ \cite{Olshanii1998,Ilg2018}, and therefore can be ignored here. 
Then, the interaction part of the  Hamiltonian is given by
\begin{equation}
 	 H_{\rs{I}}=
  	\frac{1}{2L} \sum_{p,k,q}V(q)a^\dagger_{p+q}a^\dagger_{k-q}a_ka_p,
\nonumber
\end{equation}
where $L$ is the quantization volume. 
From the microscopic Hamiltonian one obtains the mean-field energy,
the single-particle Bogoliubov excitation spectrum, as well as the leading beyond-mean-field correction within standard Bogoliubov theory.
However, it is well established that for dipolar quantum gases, 
the beyond-mean-field correction plays a crucial role in stabilizing the quantum droplets 
and needs to be included when describing the excitation spectrum.
So far, the analysis is mainly based on numerical studies of the extended Gross-Pitaevskii equation, 
where the beyond mean-field term is included within local-density approximation \cite{Roccuzzo2019,Smith2023}. 
In analogy, we add a term $H_{\rs{LHY}}$ to the Hamiltonian such that Bogoliubov theory on this effective Hamiltonian 
properly accounts for the low-energy excitations within Bogoliubov theory; 
this method is equivalent to studying the excitation spectrum within the extended Gross-Pitaevskii equation, but more suitable  for our analytical study.

The beyond-mean-field correction for a three-dimensional dipolar Bose-Einstein condensate has been determined in Refs.~\cite{Lima2011,Lima2012}. 
The energy density takes the form 
\begin{align}
  	u_{\rs{LHY}}=\frac{256\sqrt{\pi}\hbar^2}{15m}(n_{\rs{3D}}a_{\rs{s}})^{5/2}\mathcal{Q}_5(\varepsilon_{\rs{dd}}),
\nonumber
\end{align}
with $\mathcal{Q}_5(\varepsilon_{\rs{dd}})=\int_0^1du (1-\varepsilon_{\rs{dd}}+3\,\varepsilon_{\rs{dd}}u^2)^{5/2}$ 
and $n_{\rs{3D}}$ the density of the homogeneous three-dimensional system. 
The function $\mathcal{Q}_5(\varepsilon_{\rs{dd}})$ accounts for the modification due to the additional dipolar interaction 
to the well-established result for contact interactions derived by Lee, Huang, and Yang (LHY) \cite{Lee1957,Lee1957a}. 
Within this derivation, one finds that the LHY correction is dominated by excitations around the momenta $1/\xi$ 
with characteristic length scale $\xi = \hbar/\sqrt{2 m n_{\rs{3D}} g}$. 
This implies that the local-density approximation is well justified, if the density varies smoothly on this characteristic scale $\xi$, i.e., $\lp \gg \xi$. 
Within local-density approximation and the using the variational wave function $\psi(x,y)$, 
we end up with the correction
\begin{align}
	\frac{E_{\rs{LHY}}}{L}=\frac{2}{5}\gamma n^{5/2}
 	&&
 	\text{where}
 	&&
 	\gamma=
 	\frac{256}{15\pi}\frac{\hbar^2}{m (\sigma\lp)^3}\mathcal{Q}_5(\varepsilon_{\rs{dd}})a_{\rs{s}}^{5/2}.
\nonumber
\end{align}
The ground-state energy including the LHY correction hence becomes
\begin{equation}
	\frac{E}{L}  = \min_{\sigma,\nu} 
	\left[
		n\, E_t(\sigma,\nu) + \frac{1}{2}  n^2 V(0) + \frac{2}{5}\gamma n^{5/2}
	\right]
	,
	\nonumber
\end{equation}
with $n$ the one-dimensional particle density $n=N/L$.
The  correction to the mean-field energy provides a correction in the chemical potential 
\begin{align}
	\mu=\frac{dE}{dN}=E_{t} + n V(0)+\gamma n^{3/2},
  \label{chempot}
\end{align}
where $N$ is the particle number. Note, that  $\sigma$ and $\nu$ are only very weakly depending on the number of particles 
and within our analysis we self-consistently ignore this  small contribution.  
Accordingly, a correction to the chemical potential affects the compressibility $\varkappa = d \mu / dn$, which gives rise to a modified sound velocity of the superfluid,
\begin{equation}
	c^2 = \frac{n \varkappa}{m} = \frac{1}{m} \left(  n V(0) +  \frac{3}{2}\gamma n^{3/2} \right). 
\label{soundvelocity}
 \nonumber
\end{equation}
The term $H_{\rs{LHY}}$ we add to the Hamiltonian is therefore determined such that it reproduces the correct ground-state energy $E$
within mean field as well as the correct sound velocity as the low-momentum limit of the excitation spectrum $\epsilon(q)$ within lowest-order Bogoliubov theory.  
The  contribution to the Hamiltonian, which fulfills these conditions, can be written as
\begin{align}
	H_{\rs{LHY}}=\frac{2}{5}\gamma\int dz \left( \Psi^\dagger(z)\Psi^{\dagger}(z)\Psi(z)\Psi(z) \right)^{5/4} 
\nonumber
\end{align}
with
\begin{align}
	\Psi(z)=\frac{1}{\sqrt{L}}\sum_q e^{i q z}a_q,
\nonumber
\end{align}
as will be demonstrated below. 
The effective Hamiltonian 
\begin{equation}
	H=H_0+H_{\rs{I}}+H_{\rs{LHY}}
\label{EffectiveHamiltonian}
\end{equation}
will allow us to determine the low-energy excitation spectrum across the phase transition from the superfluid to the supersolid. 
The validity of our approach is limited to momentum $q \ll1 /\xi$  and energies  $\epsilon(q) \ll  \mu$ such that the local-density treatment for the term $H_{\rs{LHY}}$ is justified. 
In addition, we require $\epsilon(q)\lesssim E_{\perp}$ in order to neglect transverse excitations.

%%%%%%%%%%%%%%%%%%%% Excitations in the superfluid phase %%%%%%%%%%%%%%%%%%%%

\section{Excitations in the superfluid phase}
\label{sec:excitationsuperfluid}
We start with the study of the excitation spectrum in the superfluid using the standard Bogoliubov procedure \cite{Bogoliubov1947}.
It is important to point out, that even in one dimension, the Bogoliubov theory provides the correct excitation spectrum in the weakly interacting regime  
as can be seen by a comparison with the exact Lieb-Liniger theory for bosons with contact interactions \cite{Lieb1963,Lieb1963a}. 
One can understand this phenomenon as locally there are still a high number of particles in the condensate, 
while quantum fluctuations only suppress the coherence between these local condensates on large distances giving rise to the 
well-established algebraic behavior \cite{Haldane1981,Petrov2000}.  
In the following, it is convenient to work in the grand canonical ensemble described by the chemical potential $\mu$
and self-consistently determine the chemical potential to find the correct particle density $n$. 
Within mean-field theory, we replace the operator $a^{\dagger}_0 = \sqrt{L n}$ by the local particle density. 
Inserting this ansatz in the grand canonical potential $\Omega$ provides, as required, the ground-state energy including the LHY correction,
\begin{equation}
	\Omega=E-\mu N,
\nonumber
\end{equation}
and we recover the relation between the particle number $n$ and the chemical potential in Eq.~(\ref{chempot}) by minimizing $\Omega$.
In the next step, we can use the standard Bogoliubov prescription to derive the excitation spectrum. 
For this purpose, we write for the bosonic field operator
\begin{align}
	\Psi(z)
	\rightarrow
	\sqrt{n}
	+
	\frac{1}{\sqrt{L}}\sum_{q\neq 0} e^{i q z}a_q
	=
	\sqrt{n}+\delta\psi(z).
\nonumber
\end{align}
Note, that within this approach with fixed chemical potential and leading-order expansion, 
we do not have to distinguish between the particle density $n$ and the condensate density $n_0$, as the difference only becomes relevant for the higher-order corrections. 
Inserting the bosonic field operator into the Hamiltonian and expanding it up to second order in the small fluctuations $\delta\psi$, 
we end up with a quadratic Hamiltonian $H_{\rs B}$ accounting for the Bogoliubov excitations
\begin{align}
	H_{\rs B}= 
	 \frac{1}{2}\sum_{q\neq 0}
	:\begin{pmatrix}
		a^\dagger_q\\
		a_{-q}
	\end{pmatrix}
	\left[ 
	\begin{pmatrix}
		\chi && 0\\
		0 && \chi
	\end{pmatrix}
	+
	\begin{pmatrix}
		\eta && \eta \\
		\eta && \eta
	\end{pmatrix}
	\right]
	\begin{pmatrix}
		a_q\\
		a_{-q}^\dagger
	\end{pmatrix}
	 :\,.
\label{eq:hamiltonianquad}
\end{align}
Here, $\colon\!\hat{O}\colon$ denotes the normal ordered operator $\hat{O}$, and we introduced the two parameters
\begin{align}
	\chi&=\epsilon_0(q)+ E_{t} +  n V(0)+\gamma n^{3/2}-\mu\,,
	\nonumber
	 \\
	 \eta&=n V(q)+\frac{3}{2}\gamma n^{3/2}.
\nonumber
\end{align}
To obtain the excitation spectrum $\epsilon(q)$, 
we diagonalize the Hamiltonian \eqref{eq:hamiltonianquad} via the Bogoliubov transformation  $a_q=u_qb_q+v_qb^\dagger_{-q}$.
The amplitudes $u_p$ and $v_p$ have to fulfill the constraint $u_p^2-v_p^2=1$ for the transformation to be canonical.
A short calculation yields the diagonal Hamiltonian for the excitation spectrum
\begin{align}
	H_{\rs B}=
	\sum_{q\neq 0} \epsilon(q) b_q^\dagger b_q,
\nonumber
\end{align}
where the Bogoliubov excitation spectrum is given by
\begin{align}
	\epsilon(q)^2=\chi^2+2\chi\eta.
\label{eq:excgeneralroton} 
\end{align}
The Bogoliubov excitation spectrum $\epsilon(q)$ depends on the chemical potential.  
However, using the correct chemical potential in Eq.~(\ref{chempot}) including the LHY correction, the excitation spectrum becomes gapless,
\begin{align}
	\epsilon(q)^2=\epsilon_0(q)^2+2\epsilon_0(q)\left[nV(q)+\frac{3}{2}\gamma n^{3/2}\right],
\label{eq:excitations}
\end{align}
as required by the famous Hugenholtz and Pines relation \cite{Hugenholtz1959}. 
At low momenta, we recover the predicted sound velocity,
\begin{align}
	\epsilon(q)\overset{q\rightarrow 0}{=}\hbar |q| c\,,
\nonumber
\end{align}
with $c$ given in Eq.~(\ref{soundvelocity}). 
Therefore, we demonstrated that our effective approach with the Hamiltonian in Eq.~(\ref{EffectiveHamiltonian}) 
is capable to reproduce the low-energy excitation spectrum within  Bogoliubov theory.

%%%%%%%%%%%%%%%%%%%% Roton Instability  %%%%%%%%%%%%%%%%%%%%
\section{Roton Instability}
\label{sec:rotoninstability}
The competition of the contact repulsion between the bosons and the attractive part of the dipole-dipole interaction 
provides a characteristic Bogoliubov excitation spectrum exhibiting a roton-like structure in the tube. 
Especially, for an increasing strength of the dipole-dipole interaction, the excitation spectrum exhibits a minimum at a finite momentum $k_{\rs{min}}$, 
which eventually can reach zero and gives rise to an instability of the superfluid.
We will briefly discuss the behavior of the excitation spectrum.  
The two different interactions in combination with the transverse trapping potential offer a high level of control on the spectrum. 
The different parameters are most conveniently expressed by the dimensionless quantities
\begin{align}
	\kappa=na_{s} \propto\frac{ \lp^2}{\xi^2},
	&&
	\varepsilon_{\rs{dd}}=\frac{a_{\rs{dd}}}{a_{\rs s}},
	&&
	\text{and}
	&&
	\lambda=\frac{a_{\rs s}}{\lp}.
\nonumber
\end{align}
Here, $\kappa$ controls the dimensionality of the system, and in our one-dimensional geometry within local-density 
approximation we require  $\kappa  \gg 1$.
In addition, the condition of a weakly interacting Bose gas requires $\lambda \ll 1$ \cite{Ilg2018}.
By tuning these three parameters, the position and energy of the roton excitation can be influenced. 
We are interested in the region, where the  superfluid  becomes unstable and transitions to the supersolid phase.
This critical point is determined by the two conditions,
\begin{align}
	\epsilon(k_{\text{r}})^2=0,
	&& \text{and} &&
	\frac{d\epsilon(q)^2}{dq}\Big|_{q=k_\text{r}}=0,
\nonumber
\end{align}
which we solve numerically. 
For our discussion, we consider a  parameter regime comparable to recent dysprosium experiments \cite{Boettcher2019,Guo2019}.
Throughout this paper we set the instability to appear at $\kappa_{\rs{c}}=11.931$ and  $\lambda_{\rs{c}}=1/200$, 
which provides the critical values $\varepsilon_{\rs{dd,c}}=1.34$,  $k_{\rs r}l_{\perp}=1.570$ and allows for a second-order phase transition (see below).
Note, that the  wave vector $k_{\rs r}$ of the roton instability for these parameters satisfies the condition of low momenta with $k_{\rs r} < 1/ \xi$.
It should also be noted, that for $\varepsilon_{\rs{dd}}>1$ the function $\mathcal{Q}_5(\varepsilon_{\rs{dd}})$  picks up a very small imaginary part.
This contribution is an unphysical artifact from local-density approximation since  a three-dimensional homogeneous dipolar gas exhibits a phonon instability 
for $\varepsilon_{\rs{dd}}>1$ \cite{Lahaye2009}.
Therefore, we drop the imaginary part in the following.
\begin{figure}[t]
  \centering
  \includegraphics[scale=1.0]{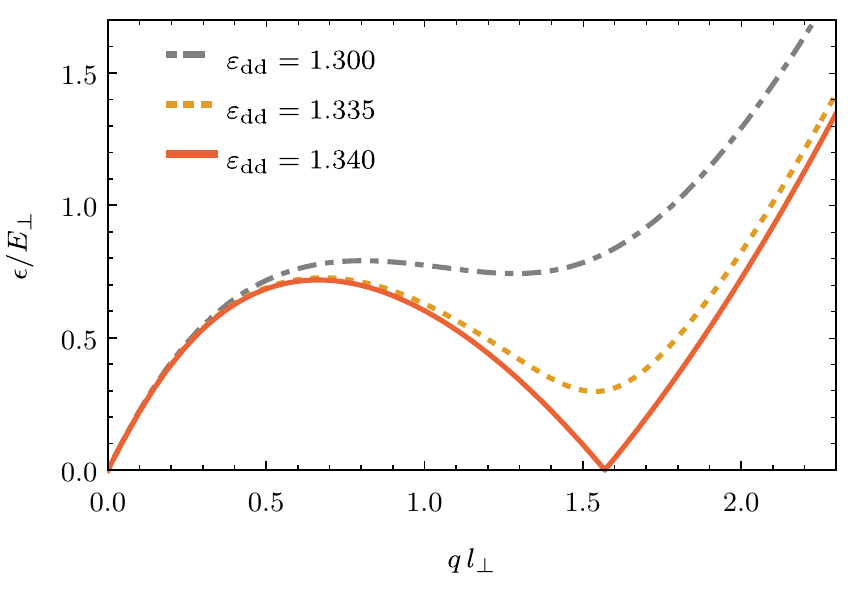}
  \caption{
	Excitation spectrum in the superfluid for different values of $\varepsilon_{\rs{dd}}$ for $\kappa_{\rs{c}}=11.931$ and $\lambda_{\rs{c}}=1/200$. 
    	By increasing $\varepsilon_{\rs{dd}}$  the spectrum develops a minimum, 
    	which eventually reaches zero energy at the critical point.
     }
  \label{fig:rotoncombined}
\end{figure}
In Fig.~\ref{fig:rotoncombined}  we compare the excitation spectrum from Eq.~\eqref{eq:excitations} for different values of $\varepsilon_{\rs{dd}}$.
We want to point out that changing $\varepsilon_{\rs{dd}}$ experimentally is achieved by tuning the scattering length $a_s$ which also affects $\kappa$ and $\lambda$.
By increasing $\varepsilon_{\rs{dd}}$ a minimum develops, which eventually reaches zero energy at the critical point $\varepsilon_{\rs{dd,c}}$.
For $\varepsilon_{\rs{dd}}>\varepsilon_{\rs{dd,c}}$ the excitation spectrum becomes imaginary close to the roton momentum indicating an instability and a breakdown of our current treatment.

%%%%%%%%%%%%%%%%%%%% Ground state in the supersolid regime  %%%%%%%%%%%%%%%%%%%%
\section{Ground state in the supersolid regime}
\label{sec:groundstatess}
The roton instability indicates the formation of a new ground state with a density modulation with  wavelength close to the corresponding roton momentum. 
Within our Bogoliubov approach, this is accounted for by the macroscopic occupation of not only the $q=0$ mode, but also the modes with $q={l k_{\rs s}}$ with $l\in \mathbb{Z}$;  
the latter give rise to a density modulation with momentum $k_{\rs{s}}$ and break the continuous translational symmetry resulting in a supersolid. 
In the following, we study first the ground state within this supersolid phase and in a next step its excitation spectrum.
The mean-field ansatz takes the form
\begin{align}
	a_{0}\rightarrow \sqrt{L n_0}
	&&
	\text{and}
	&&
	a_{\pm l k_{\rs s}}\rightarrow \frac{\Delta_l}{2}\sqrt{L n_0}\:e^{\pm i l \varphi},
\nonumber
\end{align}
with the order parameters $\Delta_l$ accounting for the solid structure. 
The bosonic field operator within mean-field theory is replaced by the condensate wave function
\begin{align}
	\Psi(z)\rightarrow  \phi(z) \equiv  \sqrt{n_0}\left( 1+ \sum_{l=1}^\infty \Delta_l\cos\left[l \: k_{\rs{s}}z+l \varphi \right]\right).
\label{eq:mfansatz}
\end{align}
We also added a phase $\varphi$ for the mean field, which illustrates the possibility to freely shift the density wave in position.  
Different to our previous treatment, the zero-momentum mode is not occupied by all particles and the total particle density is given by
\begin{align}
	n= \frac{1}{L}\int dz |\phi(z)|^2=n_0\left(1+\sum_{l=1}^{\infty}\frac{\Delta_l^2}{2}\right).
\label{eq:condnumber}
\end{align}
Note that for only one order parameter, the ansatz in Eq.~\eqref{eq:mfansatz}  reduces to the cosine-modulated ansatz used in \cite{Blakie2020a}.
Inserting the mean-field wave function into the  effective Hamiltonian $H$, the energy depends on the order parameters $\Delta_l$ as well as the wave vector $k_{\rs{s}}$, i.e.,
$E(\bm{\Delta}=(\Delta_1,\Delta_2,\dots, k_{\rs s}))$,
which  is  conveniently expressed by
\begin{align}
	E(\bm{\Delta})=&-\int dz \,\phi^*(z)\frac{\hbar^2\nabla^2}{2m}\phi(z) + NE_{t}(\sigma,\nu)
\nonumber\\
	&+
  	\frac{1}{2}\int dz \,dz'\, V(z-z')|\phi(z')|^2|\phi(z)|^2
\label{eq:energygen}
 	 \\
  	&+
  	\frac{2}{5}\gamma\int dz\,|\phi(z)|^5,
\nonumber
\end{align}
where $V(z)$ is the effective 1D interaction potential in real space.
The energy has been evaluated analytically for up to four order parameters (see Appendix \ref{app:groundstate}), 
while including more order parameters will not drastically improve our ansatz (see below). 
Note that the energy also still depends on the transverse variational parameters $\sigma$ and $\nu$ but not on the phase $\varphi$.
Varying $\varphi$ results in displacing the entire modulated state within the tube, and accounts for the broken continuous translation symmetry, 
which will give rise to an additional Goldstone mode \cite{Nielsen1976}. 
Without loss of generality, we set $\varphi=0$ in the following. 
The ground state is obtained by minimizing the energy $E(\bm{\Delta})$ with respect to $\bm{\Delta}$,  $\sigma$, and  $\nu$,
under the constraint of a fixed particle number $n$ in Eq.~(\ref{eq:condnumber}), resulting in the parameters $\bm{\Delta}_{\rs{gs}}$.
\begin{figure}[htpb]
  \centering
  \includegraphics{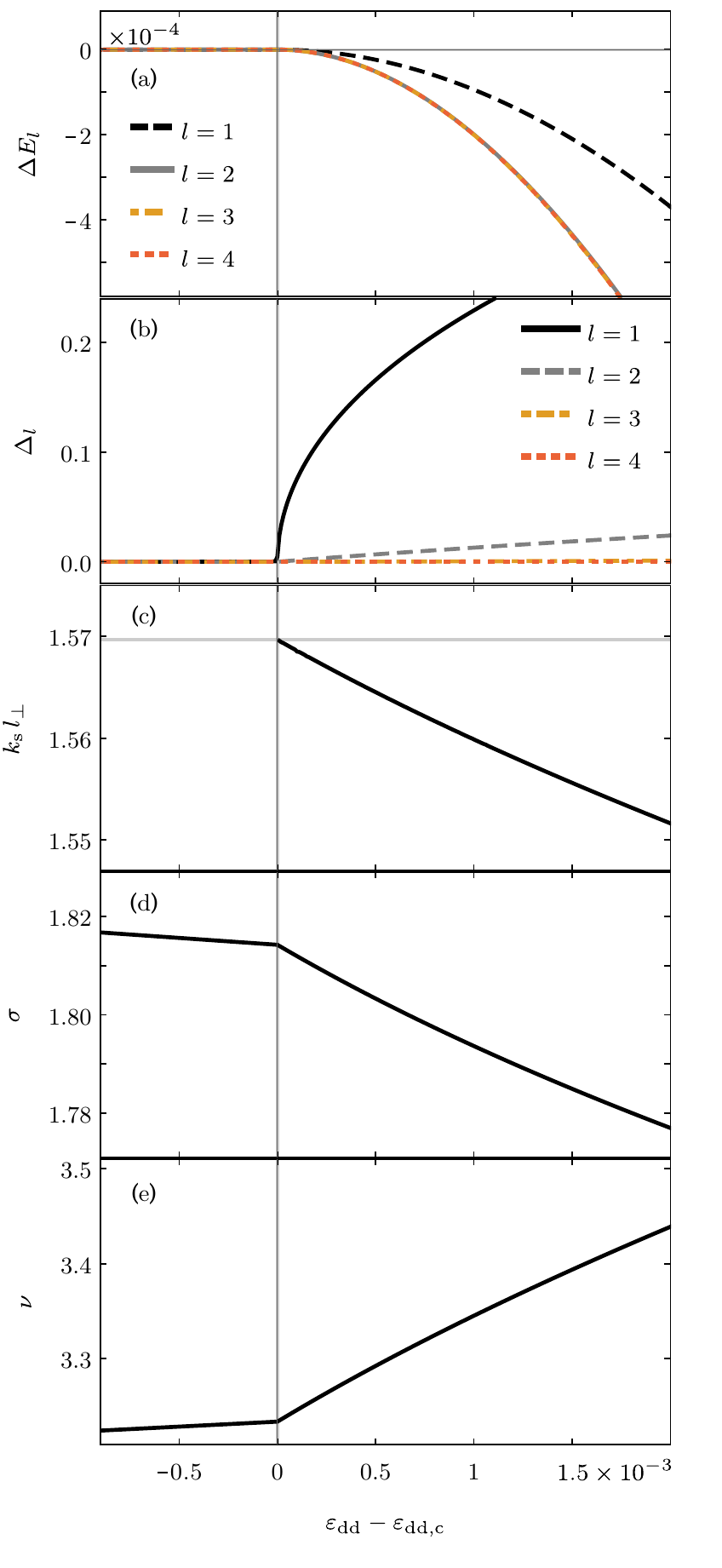}
  \caption{
  	Ground state-parameters across the superfluid to supersolid phase transition as a function of $\varepsilon_{\rs{dd}}$ for $\kappa_c=11.931$ and $\lambda_{c}=1/200$.
	(a) Energy difference per particle $\Delta E_l=[E(\bm{\Delta}_{\rs{gs}})-E(0)]/(E_\perp N)$ between the ground state of Eq.\eqref{eq:energygen} 
  	and the superfluid when including $l=1...4$ order parameters.
  	(b) The four order parameters used to obtain $\Delta E_4$ as a function of $\varepsilon_{\rs{dd}}$.
  	(c) Wave vector $k_{\rs s}$ for the density modulation in the supersolid state. 
  	The gray horizontal line shows the roton momentum $k_{\rs r}$.
  	(d) Transverse width $\sigma$ and (e) transverse anisotropy $\nu$ of the ground-state wave function.
  }
  \label{fig:dE}
\end{figure}
The superfluid state is then given by $\Delta_l=0$, while the phase transition into the supersolid phase is characterized by a finite $\Delta_l \neq 0$.
The chemical potential is determined by
\begin{equation}
	\mu =\frac{1}{L} \frac{d E(\bm{\Delta}_{\rs{gs}})}{d n}. 
\nonumber
\end{equation}
Note, that  $\bm{\Delta}_{\rs{gs}}$ also depends on the density $n$, but in analogy to the treatment in the superfluid phase we neglect the weak density dependence of the transverse degrees of freedom $\sigma$ and $\nu$.

Since the phase diagram of this model has been studied in \cite{Blakie2020a} and 
an accurate phase diagram of the microscopic parameters would require to include the transverse degrees of freedom not only variationally \cite{Smith2023},
we focus on the parameters $\kappa_{\rs{c}}$ and $\lambda_{\rs{c}}$ that allow for a second-order phase transition and investigate the stability in the thermodynamic limit.
In Fig.~\ref{fig:dE}(a), we show the energy difference per particle, $\Delta E_l=[E(\bm{\Delta}_{\rs{gs}})-E(0)]/(E_{\perp} N)$, 
between the minimized energy in Eq.~\eqref{eq:energygen} when including $l$ order parameter and the energy of the superfluid, $E(\bm{\Delta}=0)$. 
For $\varepsilon_{\rs{dd}}<\varepsilon_{\rs{dd,c}}$ the energy difference vanishes and the superfluid is the ground state of the system. 
Increasing $\varepsilon_{\rs{dd}}$ beyond $\varepsilon_{\rs{dd,c}}$, we find a continuous phase transition into the supersolid phase. 
While a single order parameter very poorly describes the ground-state energy across the phase transition (black dashed line) 
the impact of more than two order parameters on the results is negligible within the studied parameter range. 
It shows that our ansatz converges fast with the number of order parameters. 
In Fig.~\ref{fig:dE}(b)  we show the four lowest order parameters $\Delta_l$, which clearly exhibit a continuous behavior consistent with a second-order phase transition. 
In addition, the density modulation at the critical point appears at the position of the roton instability $k_{\rs r}$ [see Fig.~\ref{fig:dE}(c)], 
but $k_{\rs s}$ is slightly lowered for increasing $\varepsilon_{\rs{dd}}$, i.e., the lattice spacing increases. 
This behavior can be understood as the side-by-side orientation of the dipoles pushes neighboring droplets further apart for an increasing dipolar strength.

For the parameters $\kappa_{\rs{c}}$ and $\lambda_{\rs{c}}$, a single order parameter does not describe the ground state accurately but predicts the correct type of phase transition,
which is not generally true.
For first-order transitions, including only a single order parameter can falsely predict a continuous transition, while including more order parameters clearly indicates a discontinuous transition
(see Appendix \ref{app:firstorder}).
Thus, we want to emphasize that a simple cosine-modulated ansatz for the supersolid can be very misleading.

%%%%%%%%%%%%%%%%%%%% Excitations in the supersolid %%%%%%%%%%%%%%%%%%%%
\section{Excitations in the supersolid}
\label{sec:excitationsss}
Next, we study the excitation spectrum within the supersolid phase and generalize the procedure introduced for the superfluid.
Since our results converge very fast with the number of order parameters, it is sufficient to include only two order parameters in the analysis.
We  again expand the field operator around the mean-field values  $\Psi(z) = \phi(z)+\delta\psi(z)$ and derive the Hamiltonian up to second order in $\delta\psi(z)$, 
which leads to a quadratic Hamiltonian in the creation and annihilation operators $a_q^{(\dagger)}$. 
Due to the broken translational symmetry in the supersolid state, the excitations are only characterized by their quasi-momentum within the first Brillouin zone 
and couple states with a momentum difference of $\pm l k_{\rs s}$.
Therefore, the excitations exhibit a behavior similar to the well-known band structure in solids; however, as we are interested in the low-energy modes, we only analyze the lowest band.
The Hamiltonian takes the form
\begin{align}
H_{\rs B}=
	\frac{1}{2}\sum_{q\in\text{1.BZ}}
	\!:\!
	\begin{pmatrix}
		\bm{a}_{+}^{\dagger}\\
		\bm{a}_{-}
	\end{pmatrix}
	\left[
	\begin{pmatrix}
		\bm{\chi} &0\\
		0 & \bm{\chi}
	\end{pmatrix}
	+
	\begin{pmatrix}
	    	\bm{\eta}&\bm{\eta}\\
	    	\bm{\eta}&\bm{\eta}
	\end{pmatrix}
	\right]
	\begin{pmatrix}
	  	\bm{a}_{+}\\
	  	\bm{a}_{-}^{\dagger}
	\end{pmatrix}
	\!:\!
\label{eq:hamiltonianquadgen}
\end{align}
where
\begin{align}
  	\bm{a}_{\pm}=
  	\begin{pmatrix}
    		a_{\pm q}\\
    		a_{\pm(q+k_{\rs{s}})}\\
    		a_{\pm(q-k_{\rs{s}})}\\
    		a_{\pm(q+2k_{\rs{s}})}\\
    		a_{\pm(q-2k_{\rs{s}})}\\
    		\vdots
  	\end{pmatrix}
  	\,,
\nonumber
\end{align}
and the matrices $\bm{\chi}$ and $\bm{\eta}$ depend on $\bm{\Delta}_{\rs{gs}}$ and the chemical potential $\mu$.
We obtain the excitation spectrum by diagonalizing the Hamiltonian in Eq.\eqref{eq:hamiltonianquadgen} via a Bogoliubov transformation 
$a_{i}=\sum_{\alpha}u_i^\alpha b_\alpha+v_i^\alpha b^{\dagger}_{-\alpha}$, where $i,\alpha\in\{q+l k_{\rs{s}},l\in\mathbb{Z}\}$.
Finding the eigenmodes $\epsilon_{\rs{s}}$ in the supersolid then reduces to finding the eigenvalues of $\bm{\chi}^2+2\bm{\chi}\bm{\eta}$,
\begin{align}
  	\det
  	\left(
  	\bm{\chi}^2+2\bm{\chi}\bm{\eta}-\epsilon_{\rs{s}}(q)^2\bm{1}
  	\right)
  	=0,
\label{eq:ev}
\end{align}
which generalizes Eq.\eqref{eq:excgeneralroton} to systems where more than one mode is macroscopically occupied.

As shown in the previous discussion,  the ground state close to the continuous phase transition is very accurately described by including two order parameters $\Delta_1$ and $\Delta_2$, 
i.e., the modes with momenta $q=0,\pm k_{\rs s}, \pm 2 k_{\rs s}$ macroscopically occupied. 
In the derivation of the low-energy excitation spectrum, we work also with this accuracy. 
This allows us to restrict the size of the vectors $\bm{a}_{\pm}$ to the 5 lowest momentum modes  $q, q \pm k_{\rs s}, q \pm 2 k_{\rs s}$ with $q$ in the first Brillouin zone, 
and  $\bm{\chi},\bm{\eta}$ reduce to  $5\times5$ matrices. 
We determine the expression for the matrices $\bm{\chi}$ and $\bm{\eta}$ analytically (see Appendix \ref{app:exc}) and calculate the eigenvalues numerically. 
\begin{figure}[t]
  \centering
  \includegraphics[scale=1.0]{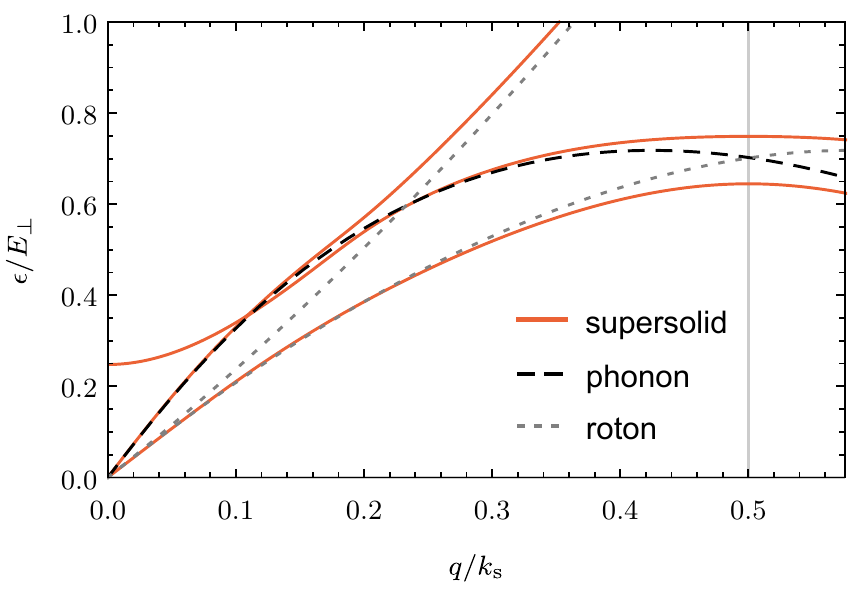}
  \caption{
    Excitation spectrum in the supersolid phase for $\varepsilon_{\rs{dd}}=1.3405$, $\lambda_{c}=1/200$, and $\kappa_{c}=11.931$.
    The red solid line shows the solution $\epsilon_{\rs{s}}$ of Eq.\eqref{eq:ev} as a function of $q\,l_\perp$.
    The gray vertical lines indicate the first Brillouin zone.
    The black and gray dashed lines show the phonon and roton branches at the phase transition, respectively, and are used as a guide to the eye.
}
  \label{fig:ev}
\end{figure}

In Fig.\ref{fig:ev} we show the excitation spectrum for $\varepsilon_{\rs{dd}}=1.3405$, close to the instability at $\varepsilon_{\rs{dd,c}}=1.34$.
The three red solid lines show the lowest eigenvalues $\epsilon_{\rs{s}}(ql_{\perp})/E_{\perp}$.
The remaining two eigenvalues contribute to higher bands and are not shown. 
We have also added the roton modes  (gray dotted lines) and  the phonon mode  (black dotted line) which 
where evaluated at the critical point; the gray vertical line indicates the first Brillouin zone.
The excitation spectrum contains two gapless modes at $q=0$, stemming from the broken $U(1)$ and translational symmetry in the supersolid, 
while the third mode shows a finite gap; the latter corresponds to the amplitude mode of the solid structure.
It is important to point out that restricting the analysis to a single order parameter $\Delta_1$ for the above parameters significantly alters the excitation spectrum.
Especially, the amplitude mode is strongly affected.
Therefore, it is crucial to accurately describe the ground state within the supersolid phase and derive the excitation spectrum with high accuracy. 

%%%%%%%%%%%%%%%%%%%%  Conclusion  %%%%%%%%%%%%%%%%%%%%
\section{Conclusion}
\label{sec:conclusion}
We present a study of the excitation spectrum of a weakly interacting gas of dipolar bosons in a tight transverse harmonic confinement across the superfluid to supersolid phase transition.
In a one-dimensional geometry, where the dipoles are aligned perpendicular to the tube, 
we introduce an effective Hamiltonian which includes beyond-mean-field effects in local-density approximation and make a variational ansatz for the transverse degrees of freedom.
The transverse confinement in combination with the dipolar interaction leads to a roton spectrum in the superfluid.
When the roton mode goes soft more than a single mode becomes macroscopically occupied and 
we adapt Bogoliubov theory by introducing an order parameter for each additional macroscopically occupied mode.
This allows us to determine the ground-state energy and the excitation spectrum across the phase transition.
For parameters comparable to current dysprosium experiments, we find that using one order parameter, 
which  corresponds to a simple cosine-modulated ansatz for the ground-state wave function in the supersolid, 
is not enough to describe the system, neither in the continuous nor in the discontinuous transition regime.
However, we show that our ansatz converges fast with the number of order parameters. 
The excitation spectrum in the supersolid regime close to a continuous transition shows no instabilities, indicating the stability in the thermodynamic limit. 
In the low-energy regime we find two gapless modes in agreement with the two broken continuous symmetries as well as a gapped amplitude mode for the solid structure.

\acknowledgments 
We thank Tilman Pfau, Chris Bühler, and Jens Hertkorn for fruitful discussions.   
This work is supported by the German Research Foundation (DFG) within FOR2247 under  Bu2247/1-2.

%%%%%%%%%%%%%%%%%%%% References  %%%%%%%%%%%%%%%%%%%%
%

%\bibliography{citations.bib}

\bibliographystyle{apsrev4-2}
%apsrev4-2.bst 2019-01-14 (MD) hand-edited version of apsrev4-1.bst
%Control: key (0)
%Control: author (8) initials jnrlst
%Control: editor formatted (1) identically to author
%Control: production of article title (0) allowed
%Control: page (0) single
%Control: year (1) truncated
%Control: production of eprint (0) enabled
%

%%%%%%%%%%%%%%%%%%%%  Appendix  %%%%%%%%%%%%%%%%%%%%
\onecolumngrid
\appendix
 
\newpage
\section{Ground-state energy}
\label{app:groundstate}
We obtain the energy in the supersolid phase as a function of $\bm{\Delta}$ by inserting the mean-field ansatz Eq.~\eqref{eq:mfansatz} into the energy functional \eqref{eq:energygen}.
For four order parameters, we evaluate the integrals analytically and obtain $E=N E_{t}+E_{\rs{kin}}+E_{\rs{int}}+E_{\rs{LHY}}$, where
\begin{align*}
	\frac{E_{\rs{kin}}}{N}=&E_{\perp}\frac{(k_{\rs{s}}\lp)^2}{4}\frac{\Delta_1^2+4\Delta_2^2+9\Delta_3^2+16\Delta_4^2}{1+\frac{\Delta_1^2}{2}+\frac{\Delta_2^2}{2}+\frac{\Delta_3^2}{2}+\frac{\Delta_4^2}{2}}\,,
	\\
	\frac{E_{\rs{int}}}{N}=&\Bigg[
       		\frac{(2+\Delta_1^2+\Delta_2^2+\Delta_3^2+\Delta_4^2)^2}{2}nV(0)
		+(\Delta_1(2+\Delta_2)+\Delta_3(\Delta_2+\Delta_4))^2nV(k_{\rs{s}})
       		\\
       		&+
       		\frac{(\Delta_1^2+2\Delta_1\Delta_3+2\Delta_2(2+\Delta_4))^2}{4}nV(2k_{\rs{s}})
       		+
       		(2\Delta_3+\Delta_1(\Delta_2+\Delta_4))^2nV(3k_{\rs{s}})
       		\\
       		&+
       		\frac{(\Delta_2^2+2\Delta_1\Delta_3+4\Delta_4)^2}{4}nV(4k_{\rs{s}})
       		+
       		(\Delta_2\Delta_3+\Delta_1\Delta_4)^2nV(5k_{\rs{s}})
       		\\
       		&+
       		\frac{(\Delta_3^2+2\Delta_2\Delta_4)^2}{4}nV(6k_{\rs{s}})
       		+
       		\Delta_3^2\Delta_4^2nV(7k_{\rs{s}})
       		+
       		\frac{\Delta_4^4}{4}nV(8k_{\rs{s}})
	\Bigg]/(2+\Delta_1^2+\Delta_2^2+\Delta_3^2+\Delta_4^2)^2.
\end{align*}
Close to the phase transition  $1+\sum_{l=1}^{4}\Delta_l \cos(lx) >0 \,\, \forall x$ and the absolute value in Eq.~\eqref{eq:energygen} can be ignored which yields
\begin{align*}
	\frac{E_{\rs{LHY}}}{N}=\gamma n^{3/2}
	\Bigg[
		&\frac{2}{5}
		+
		\frac{1}{8} 
		\Big(
		\Delta_1^4 (6 + 4 \Delta_2 + \Delta_4) + 4 \Delta_1^3 \Delta_3 (2 + 3 \Delta_2 + 3 \Delta_4) 
		\\
		&+
  		2 \Delta_1^2 \big(8 + 3 \Delta_2^3 + 12 \Delta_4^2 + 6 \Delta_2^2 (2 + \Delta_4) + 3 \Delta_3^2 (4 + \Delta_4)
		+
      		6 \Delta_2 (2 + \Delta_3^2 + \Delta_4 (2 + \Delta_4))\big) 
		\\
		&+ 
   		12 \Delta_1 \Delta_3 \big(\Delta_2^3 + 2 \Delta_2^2 (1 + \Delta_4) + \Delta_4 (4 + \Delta_3^2 + \Delta_4^2)
		+
      		\Delta_2 (4 + \Delta_3^2 + 2 \Delta_4 (2 + \Delta_4))\big) 
		\\
		&+
   		2 (\Delta_2^3 \Delta_3^2 + 3 \Delta_3^4 + 8 \Delta_4^2 + 3 \Delta_4^4 + 3 \Delta_2 \Delta_3^2 \Delta_4 (4 + \Delta_4) 
		+ \Delta_2^4 (3 + 2 \Delta_4) + 4 \Delta_3^2 (2 + 3 \Delta_4^2) )
		\\
		&+ 
      		2 (\Delta_2^2 (8 + 6 \Delta_3^2 (2 + \Delta_4) + 3 \Delta_4 (2 + \Delta_4)^2))
      		\Big)
	\Bigg]/\Big(1+\frac{\Delta_1^2}{2}+\frac{\Delta_2^2}{2}+\frac{\Delta_3^2}{2}+\frac{\Delta_4^2}{2}\Big)^{5/2}.
\end{align*}
The ground state is then obtained by minimizing the energy with respect to $\Delta_{1},\dots,\Delta_{4}$, $k_{\rs{s}}$,  $\sigma$ and $\nu$.
\section{First-order Transition}
\label{app:firstorder}
\begin{figure}[htpb]
	\centering
	\includegraphics{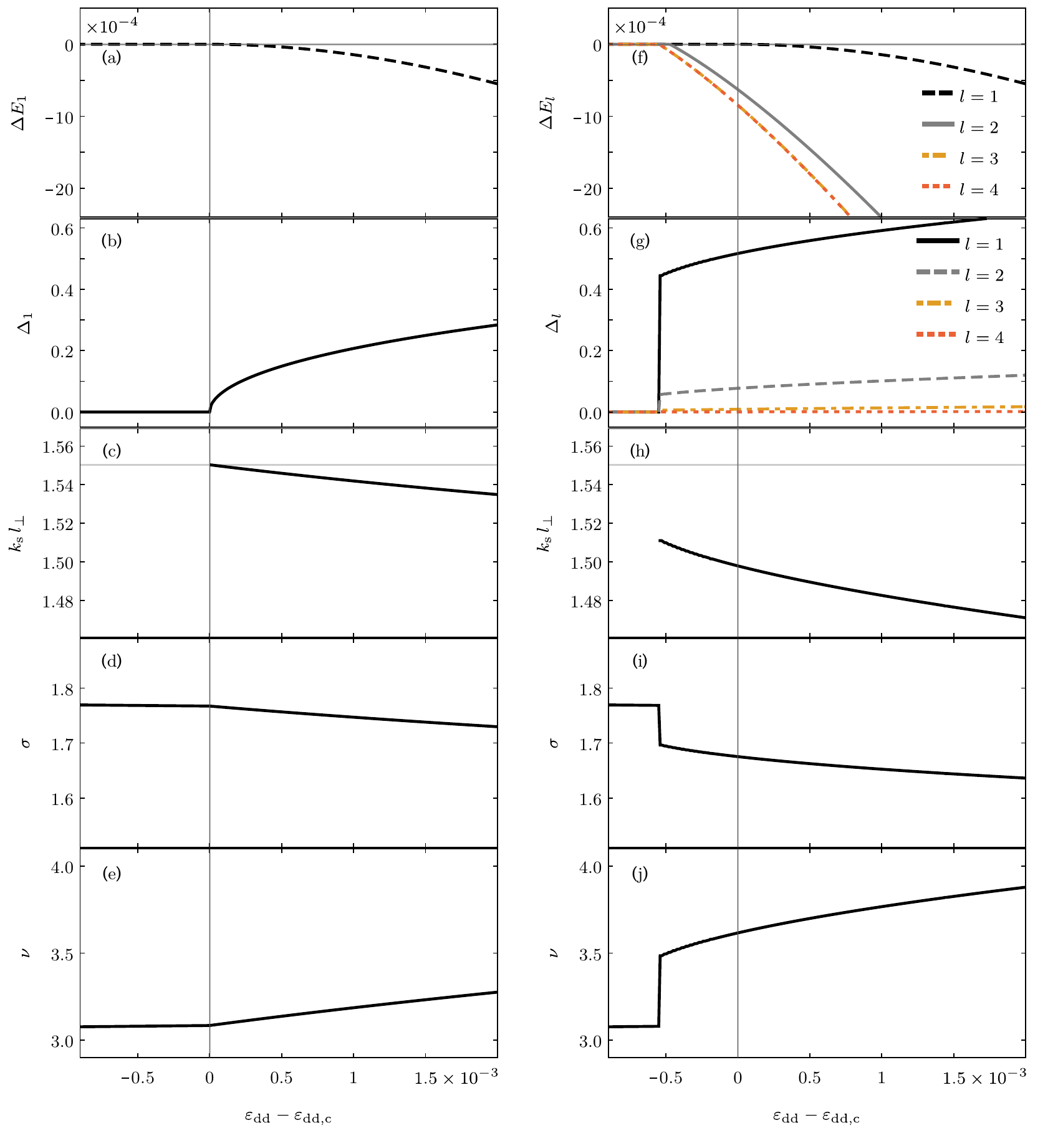}
	\caption{
		Ground-state parameters across the superfluid to supersolid phase transition as a function of $\varepsilon_{\rs{dd}}$ for $\kappa_{\rs{c}}=9.982$ and $\lambda_{\rs{c}}=1/220$.
		For (a)-(e) we only include a single order parameter. 
		(a) Energy difference per particle $\Delta E_{1}$. 
		(b) Order parameter used to obtain $\Delta E_{1}$ as a function of $\varepsilon_{\rs{dd}}$.
		(c) Wave vector $k_{\rs{s}}$ for the density modulation of the supersolid state.
		(d) Transverse width $\sigma$ and (e) transverse anisotropy $\nu$ of the ground-state wave function.
		(f) Energy difference per particle $\Delta E_{l}$ when including $l=1\dots4$ order parameters.
		For (g)-(j) we include four order parameters to obtain the system parameters' analog to (b)-(e).
		}
	\label{fig:gs1o}
\end{figure}

In addition to the continuous transition discussed in the main text, we briefly want to comment on first-order transitions within our approach.
By fixing the critical values to $\kappa_{\rs{c}}=9.982$ and $\lambda_{\rs{c}}=1/220$, the roton instability again appears at $\varepsilon_{\rs{dd,c}}=1.34$, 
however, since $\kappa_{\rs{c}}$ and $\lambda_{\rs{c}}$ are smaller compared to the values chosen in the main text ($\kappa_{\rs{c}}=11.931$, $\lambda_{\rs{c}}=1/200$)  
the term $\lambda\,\kappa^{3/2}\sim H_{\rs{LHY}}/N$ is too small and the transition is of first order.
This becomes apparent in Fig.\ref{fig:gs1o} where we show the system parameter across the phase transition.
For Figs.~\ref{fig:gs1o}(a)-(e), we only include a single order parameter in our approach, which corresponds to a simple cosine-modulated ansatz.
The energy difference per particle $\Delta E_{1}$ in Fig.~\ref{fig:gs1o}(a), the order parameter $\Delta_{1}$ in Fig~\ref{fig:gs1o}(b), the modulation $k_{\rs{s}}$ in Fig.~\ref{fig:gs1o}(c),
as well as the transverse width $\sigma$ in Fig.~\ref{fig:gs1o}(d) and the transverse anisotropy $\nu$ in Fig.~\ref{fig:gs1o}(e) all indicate a second-order phase transition, 
analogously to the parameter regime in the main text.
Including more order parameter, the discussion changes drastically.
By including more order parameters we find that the ground-state energy can be lowered even for $\varepsilon_{\rs{dd}}<\varepsilon_{\rs{dd,c}}$ [see Fig.~\ref{fig:gs1o}(f)] 
indicating a first-order phase transition.
For Figs.~\ref{fig:gs1o}(g)-(j) we show the system parameter when including four order parameters. 
In Fig.~\ref{fig:gs1o}(g), the  lowest order parameters $\Delta_{l}$ show a jump to a finite value at $\varepsilon_{\rs{dd}}\approx 1.3395$.
This discontinuous behavior also shows in the width in Fig.~\ref{fig:dE}(i) and anisotropy in Fig.~\ref{fig:dE}(j) of the ground-state wave function.
The modulation of the ground state never coincides with the roton momentum [see Fig.~\ref{fig:dE}(h)].
The previous discussion shows that a simple cosine-modulated ansatz can be very misleading for characterizing the properties of the ground state.
\section{Excitations}
\label{app:exc}
We calculate the matrices $\bm{\chi}$ and $\bm{\eta}$ by including two order parameters and expanding the Hamiltonian \ref{EffectiveHamiltonian} up to quadratic order in 
the creation and annihilation operators.
We obtain 
\begin{align*}
	\bm{\chi}=\bm{x}+\frac{1}{2}\frac{\gamma n^{3/2}}{(1+\frac{\Delta_1^2}{2}+\frac{\Delta_2^2}{2})^{3/2}}\bm{\Lambda} 
	&&\text{and} &&
	\bm{\eta}=\frac{\bm{h}}{1+\frac{\Delta_1^2}{2}+\frac{\Delta_2^2}{2}}+\frac{3}{4}\frac{\gamma n^{3/2}}{(1+\frac{\Delta_1^2}{2}+\frac{\Delta_2^2}{2})^{3/2}}\bm{\Lambda}
\end{align*}
where  $\bm{x}$, $\bm{h}$, and $\bm{\Lambda}$ are symmetric $5\times 5 $ matrices with entries
\begin{align*}
	x_{11}&=\epsilon_0(q)-\mu+E_{t}+n V(0),
	\\
	x_{22}&=\epsilon_0(q+k_{\rs{s}})-\mu+E_{t}+n V(0),
	\\
	x_{33}&=\epsilon_0(q-k_{\rs{s}})-\mu+E_{t}+n V(0),
	\\
	x_{44}&=\epsilon_0(q+2k_{\rs{s}})-\mu+E_{t}+n V(0),
	\\
	x_{55}&=\epsilon_0(q-2k_{\rs{s}})-\mu+E_{t}+n V(0),
	\\
	x_{12}&=x_{13}=x_{24}=x_{35}=\frac{\Delta _1 \left(\frac{\Delta _2}{2}+1\right) nV\left(k_s\right)}{1+\frac{\Delta _1^2}{2}+\frac{\Delta _2^2}{2}},
	\\
	x_{14}&=x_{15}=x_{23}=\frac{\left(\frac{\Delta _1^2}{4}+\Delta _2\right) nV\left(2 k_s\right)}{1+\frac{\Delta _1^2}{2}+\frac{\Delta _2^2}{2}},
	\\
	x_{25}&=x_{34}=\frac{\Delta _1 \Delta _2 nV\left(3 k_s\right)}{2 \left(1+\frac{\Delta _1^2}{2}+\frac{\Delta _2^2}{2}\right)},
	\\
	x_{45}&=\frac{\Delta _2^2 nV\left(4 k_s \right)}{4 \left(1+\frac{\Delta _1^2}{2}+\frac{\Delta _2^2}{2}\right)},
	\\
	\Lambda_{i,i}&=2 \left(\frac{3}{4} \Delta _2 \Delta _1^2+\frac{3 \Delta _1^2}{2}+\frac{3 \Delta _2^2}{2}+1\right),
	\\
	\Lambda_{12}&=\Lambda_{13}=\Lambda_{24}=\Lambda_{35}=\frac{3 \Delta _1^3}{4}+\frac{3}{2} \Delta _2^2 \Delta _1+3 \Delta _2 \Delta _1+3 \Delta _1,
	\\
	\Lambda_{14}&=\Lambda_{15}=\Lambda_{23}=\frac{3 \Delta _2^3}{4}+\frac{3}{2} \Delta _1^2 \Delta _2+3 \Delta _2+\frac{3 \Delta _1^2}{2},
	\\
	\Lambda_{25}&=\Lambda_{34}=\frac{\Delta _1^3}{4}+\frac{3}{4} \Delta _2^2 \Delta _1+3 \Delta _2 \Delta _1,
	\\
	\Lambda_{45}&=\frac{3}{4} \Delta _2 \Delta _1^2+\frac{3 \Delta _2^2}{2},
	\\
	h_{11}&=\frac{\Delta _1^2}{4}  \left[nV\left(q-k_s\right)+nV\left(k_s+q\right)\right]+\frac{\Delta _2^2}{4}  \left[nV\left(q-2 k_s\right)+nV\left(2k_s+q \right)\right]+nV\left(q\right),
	\\
	h_{22}&=\frac{\Delta _1^2}{4}  \left[nV\left(2 k_s+q\right)+nV\left(q\right)\right]+\frac{\Delta _2^2}{4}  \left[nV\left(q-k_s\right)+nV\left(3 k_s+q\right)\right]+nV\left(k_s+q\right),
	\\
	h_{33}&=\frac{\Delta _1^2}{4}  \left[nV\left(q-2 k_s\right)+nV\left(q\right)\right]+\frac{\Delta _2^2}{4}  \left[nV\left(q-3 k_s\right)+nV\left(k_s+q\right)\right]+nV\left(q-k_s\right),
	\\
	h_{44}&=\frac{\Delta _1^2}{4}  \left[nV\left(k_s+q\right)+nV\left(3 k_s+q\right)\right]+\frac{\Delta _2^2}{4}  \left[nV\left(4 k_s+q\right)+nV\left(q\right)\right]+nV\left(2 k_s+q\right),
	\\
	h_{55}&=\frac{\Delta _1^2}{4}  \left[nV\left(q-3 k_s \right)+nV\left(q-k_s \right)\right]+\frac{\Delta _2^2}{4}  \left[nV\left(q-4 k_s\right)+nV\left(q\right)\right]+nV\left(q-2 k_s\right),
	\\
	h_{12}&=\frac{\Delta _1}{2}  \left[nV\left(k_s+q\right)+nV\left(q\right)\right]+\frac{\Delta _1 \Delta _2}{4}  \left[nV\left(q-k_s\right)+nV\left(2k_s+q\right)\right],
	\\
	h_{13}&=\frac{\Delta _1}{2}  \left[nV\left(q-k_s \right)+nV\left(q\right)\right]+\frac{\Delta _1 \Delta _2}{4}  \left[nV\left(q-2 k_s\right)+nV\left(k_s+q\right)\right],
	\\
	h_{14}&=\frac{\Delta _1^2}{4}  nV\left(k_s+q\right)+\frac{\Delta _2}{2}  \left[nV\left(2 k_s+q\right)+nV\left(q\right)\right],
	\\
	h_{15}&=\frac{\Delta _1^2}{4}  nV\left(q-k_s\right)+\frac{\Delta _2}{2}  \left[nV\left(q-2 k_s\right)+nV\left(q\right)\right],
	\\
	h_{23}&=\frac{\Delta _2}{2}  \left[nV\left(q-k_s\right)+nV\left(k_s+q\right)\right]+\frac{\Delta _1^2}{4}  nV\left(q\right),
	\\
	h_{24}&=\frac{\Delta _1}{2}  \left[nV\left(k_s+q\right)+nV\left(2 k_s+q\right)\right]+\frac{\Delta _1 \Delta _2}{4}  \left[nV\left(3 k_s+q\right)+nV\left(q\right)\right],
	\\
	h_{25}&=\frac{\Delta _1 \Delta _2}{4} \left[nV\left(q-k_s\right)+nV\left(q\right)\right],
	\\
	h_{34}&=\frac{\Delta _1 \Delta _2}{4} \left[nV\left(k_s+q\right)+nV\left(q\right)\right],
	\\
	h_{35}&=\frac{\Delta _1 \Delta _2}{4}  \left[nV\left(q-3 k_s\right)+nV\left(q\right)\right]+\frac{\Delta _1}{2}  \left[nV\left(q-2 k_s\right)+nV\left(q-k_s\right)\right],
	\\
	h_{45}&=\frac{\Delta _2^2}{4} nV\left(q\right).
\end{align*}

\end{document}